\documentclass[pdflatex,sn-mathphys-num]{sn-jnl}% Math and Physical Sciences Numbered Reference Style
\usepackage{graphicx}%
\usepackage{multirow}%
\usepackage{amsmath,amssymb,amsfonts}%
\usepackage{amsthm}%
\usepackage{mathrsfs}%
\usepackage[title]{appendix}%
\usepackage{xcolor}%
\usepackage{textcomp}%
\usepackage{manyfoot}%
\usepackage{booktabs}%
\usepackage{algorithm}%
\usepackage{algorithmicx}%
\usepackage{algpseudocode}%
\usepackage{listings}%
\usepackage{rotating} % dans le préambule
\usepackage{pdflscape} % dans le préambule

\theoremstyle{thmstyleone}%
\theoremstyle{thmstyletwo}%

\theoremstyle{thmstylethree}%

\begin{document}

\title[Article Title]{
Probing the first generations of massive stars through fluorine in CEMP-no stars
}

%%=============================================================%%
%% GivenName	-> \fnm{Joergen W.}
%% Particle	-> \spfx{van der} -> surname prefix
%% FamilyName	-> \sur{Ploeg}
%% Suffix	-> \sfx{IV}
%% \author*[1,2]{\fnm{Joergen W.} \spfx{van der} \sur{Ploeg} 
%%  \sfx{IV}}\email{iauthor@gmail.com}
%%=============================================================%%

\author*[1,2]{\fnm{Arthur} \sur{Choplin}}\email{arthur.choplin@ulb.be}

\author[3]{\fnm{Georges} \sur{Meynet}}\email{georges.meynet@unige.ch}
%\equalcont{These authors contributed equally to this work.}

%\author[1,2]{\fnm{Third} \sur{Author}}\email{iiiauthor@gmail.com}
%\equalcont{These authors contributed equally to this work.}

\affil*[1]{\orgdiv{Institut d'Astronomie et d'Astrophysique}, \orgname{Universit\'e Libre de Bruxelle}, \orgaddress{\street{CP 226}, \city{Brussels}, \postcode{1050}, \country{Belgium}}}

\affil[2]{\orgdiv{BLU-ULB, Brussels Laboratory of the Universe, Brussels, Belgium}}

\affil[3]{\orgdiv{Department of Astronomy}, \orgname{ University of Geneva}, \orgaddress{\street{Chemin Pegasi 51}, \city{Versoix}, \postcode{1290}, \state{Geneva}, \country{Switzerland}}}

%%==================================%%
%% Sample for unstructured abstract %%
%%==================================%%

%\abstract{}

%%================================%%
%% Sample for structured abstract %%
%%================================%%

\abstract{\textbf{Purpose:} 
We investigate whether the first discovered fluorine-rich CEMP-no star, CS~29498$-$043, can be explained by a very metal-poor rotating massive star.

\textbf{Methods:} 
We consider single rotating stellar models of 20 $M_{\odot}$ at a metallicity of $Z = 10^{-5}$, exploring initial rotation rates from $\upsilon_{\rm ini}/\upsilon_{\rm crit} = 0$ to $0.7$ in increments of $0.1$ ($0<\upsilon_{\rm ini}<644$~km~s$^{-1}$).
 
\textbf{Results:} 
Rotational mixing enhances the production of light elements in the H--He layers, including fluorine. The ejected material can be nitrogen-rich without being fluorine-rich, whereas fluorine-rich ejecta are always predicted to be nitrogen-rich. 
The model providing the best fit to the abundances of CS~29498$-$043 is the $\upsilon_{\rm ini}/\upsilon_{\rm crit} = 0.6$ model ($\upsilon_{\rm ini} = 547$~km~s$^{-1}$), which reproduces C, N, O, Na, Mg, and Al within the observational uncertainties. However, the predicted [F/Fe]~$=2.8$ exceeds the observed value of [F/Fe]~$=2.0 \pm 0.4$.  
By simultaneously varying the $^{15}$N($\alpha,\gamma$)$^{19}$F and $^{19}$F($\alpha,p$)$^{22}$Ne reaction rates within their acceptable ranges, the [F/Fe] ratio in the $\upsilon_{\rm ini}/\upsilon_{\rm crit} = 0.6$ model can be reduced to 2.2, providing a plausible solution to the abundance pattern of CS~29498$-$043.

\textbf{Conclusion:} 
Our results support the hypothesis that fluorine-rich CEMP-no stars may originate from material enriched by a single, metal-poor, rotating massive star. A potential observational test of this scenario may be to check whether the nitrogen and fluorine abundances observed at the surface of CEMP-no stars are correlated.
}


\maketitle

\section{Introduction}\label{sec1}

While distant galaxies offer fascinating glimpses into the early Universe and possibly the first generations of stars, another valuable window is provided by halo stars in our own Galaxy that exhibit extremely low iron abundances \cite{2015ARAA, bonifacio25}. These stars are thought to be composed of interstellar material enriched by only one or two massive stars from a preceding generation. They also typically show strong carbon enhancements, leading to the definition of Carbon-Enhanced Metal-Poor (CEMP) stars\footnote{A better naming would have been Iron, rather than Metal poor, since those stars show large abundance of what in astrophysics is labeled under the term metal. } \cite{beers05}. 

As is well known, CEMP stars can be classified into several categories based on the nature and abundances of their trans-iron elements \citep[see, e.g.,][]{beers05, Frebel2018}. CEMP-no stars constitute a subclass of CEMP stars characterized by the absence of enhancements in trans-iron elements. Interestingly, these objects exhibit the lowest iron abundances observed. Such extremely low iron content is consistent with formation from a protostellar cloud enriched by only a few (perhaps even a single) massive star from an earlier generation. This progenitor was most likely massive enough that, over its nuclear lifetime, there was insufficient time to produce a significant increase in overall metallicity. 
Among CEMP-no stars, some display not only large [C/Fe] ratios\footnote{[X/Y]~$= \log_{10}(N_{\rm X} / N_{\rm Y})_{\star} - \log_{10}(N_{\rm X} / N_{\rm Y})_{\odot}$ with $N_{\rm X}$ and $N_{\rm Y}$ the number density of elements X and Y in the observed star and in the Sun.} but also high [N/Fe] and [O/Fe] ratios, along with enhanced abundances of other elements such as Na and Mg. These stars are of low mass; consequently, their lifetimes exceed the time elapsed between their birth in the very early Universe and the present day.
The CEMP-no subclass can be further divided into distinct groups according to their [Fe/H] ratios and absolute carbon abundances\footnote{A(C)~$= \log\epsilon$(C)~$= \log_{10} (N_{\rm C} / N_{\rm H}) + 12$.} A(C) \cite{yoon16, yoon19, lee25}.

Many different models have been proposed to explain the peculiar abundance patterns of CEMP-no stars. The vast majority of such models assume that the observed surface abundances require an external source, rather than resulting from internal processes within the CEMP-no stars themselves. Some CEMP-no stars are still on the main sequence, making it impossible for them to produce carbon- or oxygen-rich material internally. 

One possible explanation is that these stars might have accreted enriched material from a more massive and evolved companion that filled its Roche lobe and/or transferred mass through stellar winds. While this scenario may apply in certain individual cases \cite{suda04, suda10, campbell10, arentsen19, komiya20, gilpons25}, it is unlikely to account for the majority of CEMP-no stars. First, because these stars are of low mass, the amount of accreted material would necessarily be small. This makes it difficult to understand why CEMP-no stars observed in the red-giant phase display carbon enhancements similar to those of main-sequence CEMP-no stars, despite the expectation that any small amount of accreted material would be strongly diluted within the extended convective envelope of red giants. Second, most CEMP-no stars appear to be single. In a sample of 24 CEMP-no stars monitored over a period of up to eight years, only four were found to belong to binary systems \cite{hansen16b}. Nevertheless, new CEMP-no binaries are being regularly discovered \cite{arentsen19, bonifacio20, aguado22, caffau25}, which may support the binary scenario for some of these stars.

Most scenarios explaining the abundances of CEMP-no stars involve one or more massive stars from a previous generation that enriched the natal cloud from which the CEMP-no star formed \cite{umeda02, limongi03, iwamoto05, meynet06, hirschi07, tominaga07b, heger10, ishigaki14, takahashi14, tominaga14, choplin17a, clarkson18, ezzeddine19, chiaki20, jeena23, hartwig23, vanni23, jiang25}. The chemical composition of that cloud would then be expected to resemble the abundances observed at the surface of the CEMP-no star -- except for light and fragile elements such as lithium, which may be altered by internal processes during the main-sequence phase. During the red-giant phase, surface abundances may also be modified by dredge-up, which brings CNO-processed material to the surface.
The material now seen at the surface of a CEMP-no star is thus interpreted as a mixture of the interstellar medium at the time of the star’s formation and the ejecta of one or more massive “source stars” belonging to a previous stellar generation. 

One of the main challenges is to account for the fact that CEMP-no stars exhibiting large overabundances of C, N, and O relative to iron require the source star’s ejecta to contain material enriched in both H- and He-burning products. At first glance, this might appear straightforward, since supernova ejecta naturally include layers processed by both H and He burning. However, it has been shown that if no internal mixing occurs within the source star during its evolution—specifically, no mixing linking the H- and He-burning regions—then it is impossible to produce simultaneously large enhancements of C, N, and O relative to iron \citep[see, e.g.,][]{meynet06bis}. The $^{12}$C/$^{13}$C ratios measured in CEMP-no stars also provide evidence for mixing between the He- and H-burning layers in their progenitor stars \cite{choplin17a, molaro23, rizzuti25}.
A plausible driver of such mixing is stellar rotation. Axial rotation induces several instabilities, most notably secular shear instability and meridional circulation. These processes enable isotopes with strong abundance gradients between the H- and He-burning zones to be transported across these regions. Rotational mixing has been shown to produce substantial amounts of primary $^{13}$C, $^{14}$N, $^{19}$F, $^{22}$Ne, and even s- and p-process elements \citep[e.g.,][and references therein]{chiappini13, maeder15a, meynet15iau, limongi18, choplin18, choplin22c}. Rotation can therefore profoundly modify the nucleosynthetic output of stars, particularly at low metallicity.  
Indeed, low-metallicity stars are more compact than their higher-metallicity counterparts, and this compactness enhances the efficiency of mixing. Since diffusion timescales scale with the square of the stellar radius divided by the diffusion coefficient, smaller radii lead to shorter mixing timescales. The diffusion coefficient itself remains broadly similar in models of the same initial mass and comparable ratios of equatorial to critical rotation velocity at the zero-age main-sequence (ZAMS) \citep{meynet02b}. 
The impact of such metal-poor massive-star models on the early evolution of galaxies has been examined by several authors \citep[e.g.][]{chiappini03, chiappini06, prantzos18}. Other studies have investigated whether these models can account for the peculiar abundance patterns of CEMP-no stars \citep{meynet10}, and even of some CEMP-s stars \citep{choplin17letter}. 
It has been proposed that the wide variety of chemical compositions observed in CEMP-no stars may result from differing degrees of mixing between the H- and He-burning zones in their progenitor massive stars \citep{maeder15a, maeder15b}.

Although fluorine is among the elements that can be produced and enhanced through rotational mixing, no CEMP-no star had, until recently, shown a detectable fluorine abundance at its surface. This absence of detections does not necessarily imply that fluorine is not present; even when it is, its abundance is notoriously difficult to determine with precision (see below). Recent progress was made by \citet{mura25}, who, for the first time, reported a fluorine abundance of [F/Fe]~$=2.0 \pm 0.4$ in the CEMP-no star CS~29498$-$043 ([Fe/H] $=-3.87$). 
The identified astrophysical sites for fluorine synthesis include asymptotic giant branch (AGB) stars \citep{forestini92, mowlavi96, lugaro04, cristallo14, vescovi21}, Wolf-Rayet (WR) stars \citep{meynet00, stancliffe05, palacios05}, rotating massive stars \citep{choplin18, limongi18, roberti24, tsiatsiou25}, the neutrino process during Type II supernovae \citep{woosley88, woosley90, woosley95}, explosive H-burning during hypernova explosions \citep{mura25}, and mergers between a helium and a carbon-oxygen white dwarf \citep{longland11}. 
Galactic chemical evolution models have highlighted the importance of rotating massive stars at low metallicity to explain the observed fluorine trends \citep{womack23, grisoni25}.
The aim of the present work is to investigate whether the chemical composition of the first detected fluorine-rich CEMP-no star, CS~29498$-$043, can be explained by metal-poor, rotating massive stars. 

This paper is structured as follows. Section~\ref{sec2} describes the physical inputs of the rotating stellar models. The nucleosynthesis predicted by these models is summarized in Sect.~\ref{sec3}, with a focus on fluorine production. Sect.~\ref{sec4} presents a comparison with the observed abundances of the fluorine-rich CEMP-no star CS~29498$-$043. Discussions and conclusions are provided in Sects.~\ref{sec5} and \ref{sec6}, respectively.

\section{Input physics}\label{sec2}

The models used in this study were originally computed by \citet{choplin19} with the Geneva stellar evolution code \citep[e.g.][]{eggenberger08}. We briefly summarize the key aspects here and refer to that work for further details.
All models assume an initial mass of $M_{\rm ini} = 20$~$M_{\odot}$, which might be characteristic of the earliest stellar generations \cite[e.g.][]{susa14}. Initial rotation rates span $\upsilon_{\rm ini}/\upsilon_{\rm crit} =$ 0 to 0.7, where $\upsilon_{\rm ini}$ is the initial velocity at ZAMS and $\upsilon_{\rm crit} = \sqrt{\tfrac{2}{3}\tfrac{GM}{R_{\rm p,c}}}$ is the critical velocity at which centrifugal force balances gravity\footnote{$G$ is the gravitational constant, $M$ the total mass, and $R_{\rm p,c}$, the radius at the pole when the star is rotating at the critical limit, taken here equal to the radius at the pole of our rotating model since the polar radius show very small variation with rotation.}. The initial metallicity is $Z = 10^{-5}$ ([Fe/H] $= -3.8$) with an $\alpha$-enhanced composition following \citet{frischknecht16} and reported in Table~\ref{table:ism}.
Mass-loss prescriptions follow \citet{vink01} for $\log T_{\rm eff} \geq 3.9$ ($\dot M \propto Z^{0.85}$) and \citet{jager88} otherwise. Rotational mixing uses $D_{\rm shear}$ from \citet{talon97} and $D_{\rm h}$ from \citet{zahn92}. Each model was evolved to the end of core oxygen burning (central $^{16}$O mass fraction $< 10^{-4}$). The main model characteristics are summarized in Table~\ref{table:1}.

  \begin{table}
\caption{Initial abundances (in mass fraction) of the stellar models.  \label{table:ism}}
\begin{tabular}{@{}lc@{}}
\hline % inserts single horizontal line
Isotope  & Mass fraction  \\
\hline % inserts single horizontal line
$^{1}$H           &            7.516e-01             \\
$^{3}$He           &            4.123-05             \\
$^{4}$He          &            2.484e-01             \\
$^{12}$C           &          1.292e-06               \\
$^{13}$C           &          4.297e-09               \\
$^{14}$N           &           1.022e-07              \\
$^{15}$N           &           4.024e-10              \\
$^{16}$O           &           6.821e-06              \\
$^{17}$O           &            3.514e-10             \\
$^{18}$O           &           2.001e-09              \\
$^{19}$F           &            8.385e-11             \\
$^{20}$Ne           &          9.205e-07               \\
$^{21}$Ne           &         7.327e-10                \\
$^{22}$Ne           &          2.354e-08              \\
$^{23}$Na           &          4.135e-09               \\
$^{24}$Mg           &         2.012e-07                \\
$^{25}$Mg           &         1.030e-08                \\
$^{26}$Mg           &         1.179e-08                \\
$^{27}$Al           &           7.694e-09              \\
$^{28}$Si           &          2.060e-07               \\
 \hline
\end{tabular}
\end{table}

This work is restricted to the investigation of the scenario developed by \citet{maeder15b}, which proposes that CEMP-no stars may be composed of H- and He-rich material from previous rotating massive stars that experienced varying degrees of internal mixing. 
Consequently, our models are assumed to undergo explosions with strong fallback, ejecting only the H- and He-rich layers above the carbon-oxygen (CO) core, which are generally not significantly altered by explosive nucleosynthesis. 

Because low- and zero-metallicity massive stars are more compact, they may experience greater fallback than their solar-metallicity counterparts \citep[e.g.,][]{woosley02}, a mechanism frequently invoked to explain the abundance patterns of the most metal-poor stars \citep[e.g.,][]{umeda02, limongi03, bonifacio03, iwamoto05}.  
Recent long-term two-dimensional supernova simulations by \citet{sykes25}, based on Pop.~III progenitors of 60 and 95~$M_{\odot}$, report extensive fallback with remnant masses typically in the range $20$--$35\,M_{\odot}$, and with mass cuts clustering near the H/He shell interface. The existence of supernova light curves compatible with such strong fallback explosions further supports this mechanism \citep{moriya10, moriya18a, moriya18b, moriya19b}.  
Furthermore, rotating stars might be more difficult to explode, as they tend to develop larger CO cores (Table~\ref{table:1}), resulting in more compact progenitors \citep{mapelli20}.

\begin{table}[h]
\caption{Characteristics of the 20~$M_{\odot}$, $Z=10^{-5}$ stellar models considered in the present work. The final masses $M_{\rm fin}$ and CO core masses $M_{\rm CO}$ (the top of the CO core is defined as the location where the $^{4}$He mass fraction rises above $10^{-2}$) are given in the two last columns.}\label{table:1}%
\begin{tabular}{@{}cccccc@{}}
\toprule
Model label             & $v_{\rm ini} / v_{\rm crit}$    & $\Omega_{\rm ini} / \Omega_{\rm crit}$           &   $v_{\rm ini}$   &    $M_{\rm fin}$ [$M_{\odot}$]     & $M_{\rm CO}$   [$M_{\odot}$]          \\
\midrule
vv0              & 0.0                         & 0      &  0                     &   19.998   & 3.89 \\  
vv1              & 0.1                         & 0.15 &  88                   &   19.998   & 4.32\\  
vv2              & 0.2                         & 0.32 &  188                 &   19.960   & 4.81\\  
vv3              & 0.3                         & 0.46 &  276                 &   19.726   & 4.89\\  
vv4              & 0.4                         & 0.69 &  364                 &   19.625   & 4.99\\  
vv5              & 0.5                         & 0.71 &  454                 &   19.189   & 5.22\\  
vv6              & 0.6                         & 0.81 &  547                 &   19.222   & 4.74 \\  
vv7              & 0.7                         & 0.90 &  644                 &   19.764   &  4.94  \\  
\bottomrule
\end{tabular}
\end{table}

\section{Nucleosynthesis with rotation}\label{sec3}

\begin{figure}[t]
\centering
\includegraphics[width=0.75\textwidth]{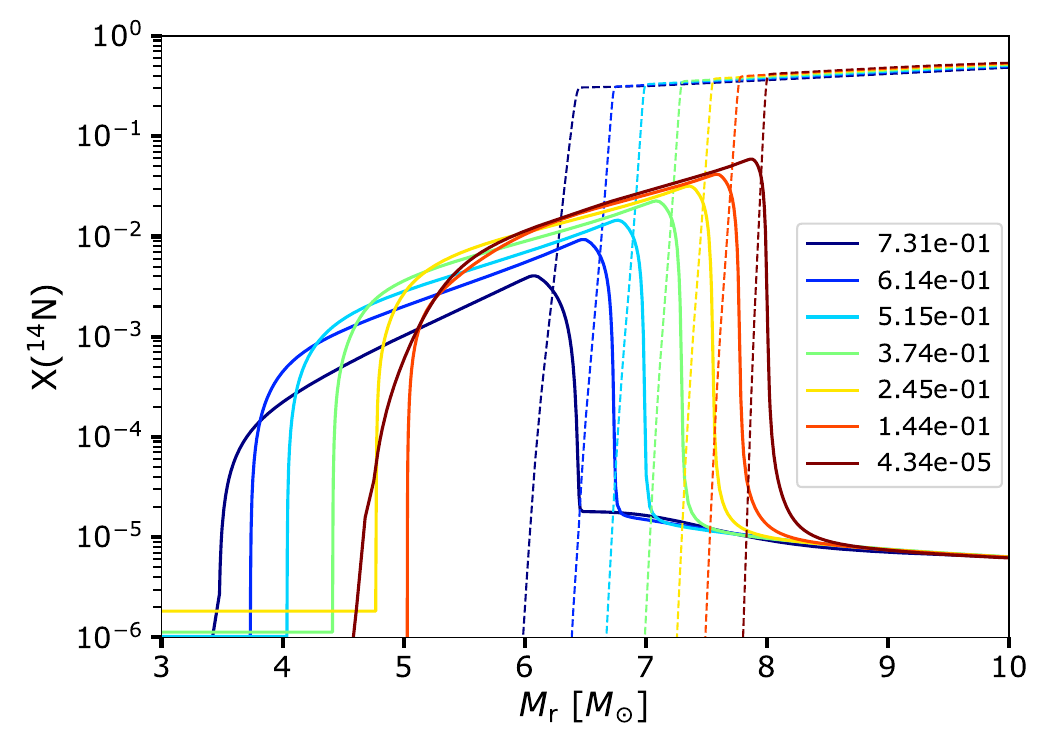}
\includegraphics[width=0.75\textwidth]{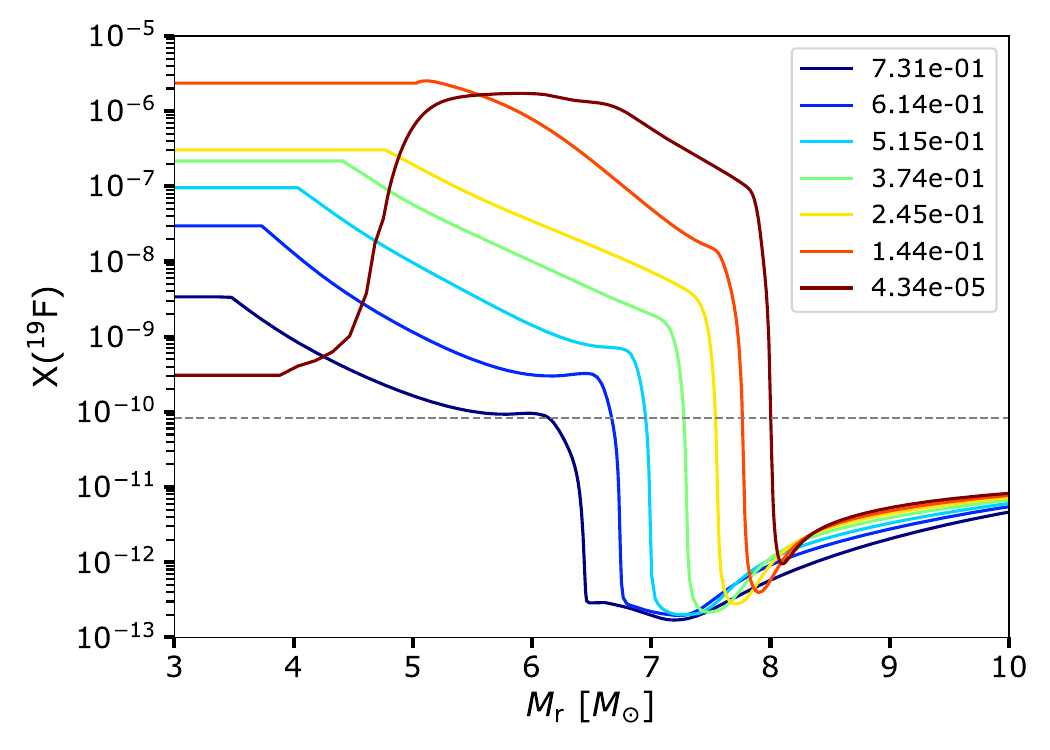}
\caption{Abundance profiles of $^{14}$N (top panel) and $^{19}$F (bottom panel) in the vv6 model during core helium burning. Different colors indicate various central helium mass fractions (ranging from 0.731 to $4.34 \times 10^{-5}$), as specified in the legend. The dashed lines in the top panel represent the $^{1}$H abundance profiles.
}
\label{fig:abhe}
\end{figure}

\begin{figure}[t]
\centering
\includegraphics[width=0.8\textwidth]{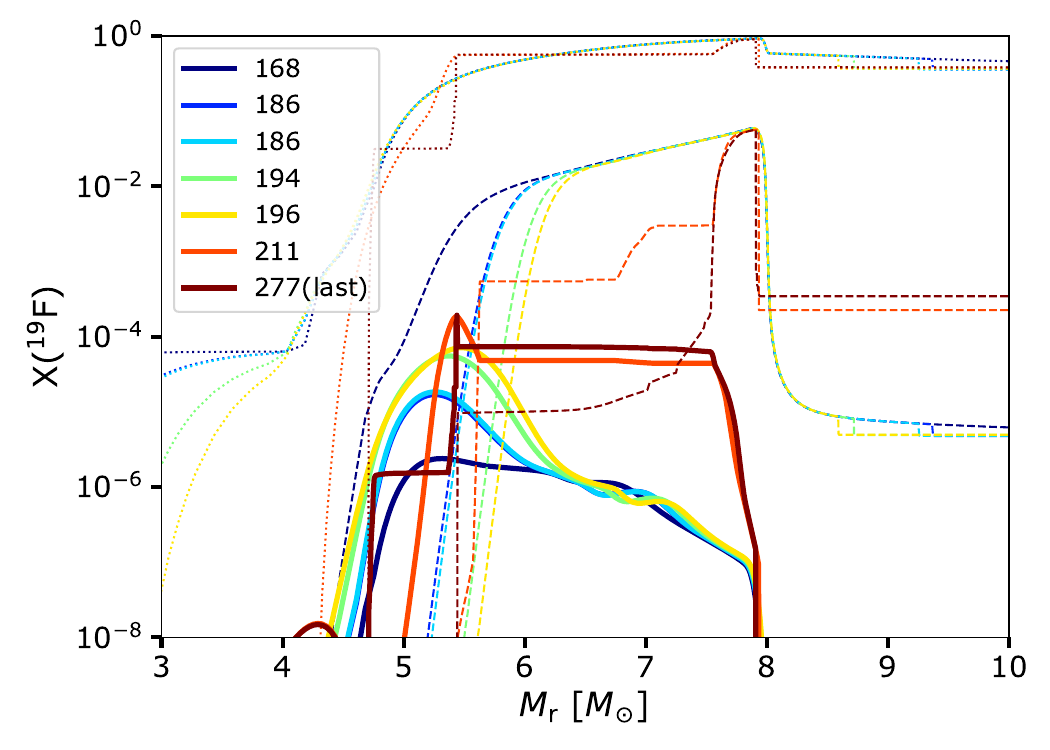}
\caption{
Abundance profiles of $^{19}$F in the vv6 model within the helium-burning shell. Different colors represent successive times. The numbers indicate the temperature (in $10^6$ K) at the peak of $^{19}$F, located near the base of the helium-burning shell. Dashed and dotted lines show the corresponding $^{14}$N and $^{4}$He profiles, respectively.
}
\label{fig:abhesh}
\end{figure}

\subsection{The back and forth mixing process}
\label{sec:backforth}
As shown, rotational mixing in massive stars triggers cyclic exchanges of material between the convective He-burning core and H-burning shell \citep[e.g.][]{meynet06, maeder15a, choplin16}. 
In the first stage, helium-burning products ($^{12}$C and $^{16}$O) are transported into the hydrogen-burning shell by rotation induced instabilities, where they enhance the CNO cycle and generate primary\footnote{“Primary” means synthesized out of the initial H and He contents, by opposition to secondary, which means synthesized out of the initial metal (elements heavier than He) content.} $^{13}$C and $^{14}$N (Fig.~\ref{fig:abhe} top panel). In turn, material from the hydrogen-burning shell, including $^{14}$N, is transported back into the helium core, where it is converted into $^{22}$Ne via $^{14}$N($\alpha,\gamma$)$^{18}$F($\beta^+\nu_e$)$^{18}$O($\alpha,\gamma$)$^{22}$Ne. Subsequent reactions lead to the production of $^{25}$Mg and $^{26}$Mg through $^{22}$Ne($\alpha,\gamma$)$^{26}$Mg and $^{22}$Ne($\alpha,n$)$^{25}$Mg, respectively\footnote{GM, One of the authors remembers explaining this mechanism and its potential impact for the nucleosynthesis of s-process elements in massive stars to Roberto Gallino in a train, coming back from Padova where we both attended a thesis presentation. This discussion had, as a result, the paper \citet{Pignatari2008}. Roberto had a passion for science and I keep a wonderful memory of this discussion.}. 
These newly formed neon and magnesium isotopes can be transported back into the hydrogen shell, activating the Ne–Na and Mg–Al chains and thus producing sodium and aluminum. 
This back and forth mixing between the hydrogen- and helium-burning zones therefore enable the synthesis of a broad range of isotopes (especially $^{13}$C, $^{14}$N, $^{22}$Ne, $^{23}$Na, $^{25,26}$Mg and $^{27}$Al), with abundances sensitive to the strength of mixing, which depends on the initial rotation rate.

\subsection{Fluorine genesis}

The sequence of processes that lead to the synthesis of F originates from Forestini et al. (1992) \cite{forestini92}.
It is primarily produced in He-burning zones, through the following reaction chains:
\begin{itemize}
\item $^{14}$N($\alpha,\gamma$)$^{18}$F($\beta^+$)$^{18}$O($p,\alpha$)$^{15}$N($\alpha,\gamma$)$^{19}$F
\item $^{14}$N($\alpha,\gamma$)$^{18}$F($n,p$)$^{18}$O($p,\alpha$)$^{15}$N($\alpha,\gamma$)$^{19}$F
\item $^{14}$N($\alpha,\gamma$)$^{18}$F($n,\alpha$)$^{15}$N($\alpha,\gamma$)$^{19}$F.
\end{itemize}
The required protons mainly originate from $^{14}$N($n,p$)$^{14}$C, while neutrons are provided by $^{13}$C($\alpha,n$)$^{16}$O (and by $^{22}$Ne($\alpha,n$)$^{25}$Mg when $T \gtrsim 250$~MK). Rotation enhances these chains by the mixing it induces between the He-burning core and the H-burning shell, which brings primary nitrogen into the He-burning core 
(Sect.~\ref{sec:backforth}), thereby efficiently activating the fluorine production channels \citep{choplin18, limongi18, tsiatsiou25}.

We focus on the fast-rotating vv6 model to follow the synthesis of $^{19}$F under efficient rotational mixing. As found in past studies \cite{choplin18, limongi18, roberti24, tsiatsiou25}, Fluorine production begins in the helium-burning core (Fig.~\ref{fig:abhe}, bottom panel), where primary $^{13}$C and $^{14}$N (Fig.~\ref{fig:abhe}, top panel) diffuse back into the core (cf. Sect.~\ref{sec:backforth}), fueling the reaction chains described above. Toward the end of helium burning (brown curve), the central temperature exceeds 300~MK, and $^{19}$F is efficiently destroyed through $^{19}$F($\alpha,p$)$^{22}$Ne. However, a significant fraction of the synthesized $^{19}$F survives (Fig.~\ref{fig:abhe}, between 5 and 8~$M_{\odot}$), largely due to rotational mixing, which transports $^{19}$F outward during core helium burning.

A second episode of fluorine production occurs in the helium-burning shell, involving the same reaction chains. During this phase, the $^{19}$F mass fraction increases from $\sim 10^{-6}$ (its abundance at the end of core helium burning) to $\sim 10^{-4}$ (Fig.~\ref{fig:abhesh}). Once again, the additional $^{13}$C and $^{14}$N generated by rotation play a crucial role in enhancing fluorine synthesis. Neutrons are mainly provided by $^{13}$C($\alpha,n$)$^{16}$O, since the shell temperature remains too low ($\lesssim 250$~MK, Fig.~\ref{fig:abhesh}) to efficiently activate $^{22}$Ne($\alpha,n$)$^{25}$Mg.

\begin{figure}[t]
\centering
\includegraphics[width=0.49\textwidth]{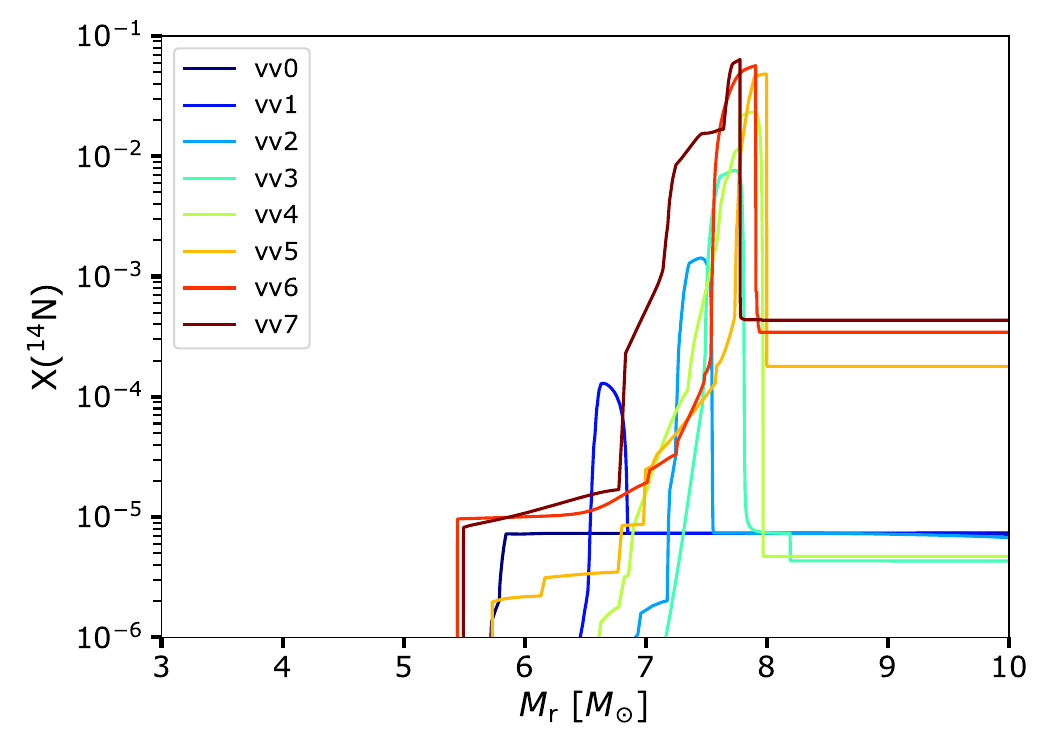}
\includegraphics[width=0.49\textwidth]{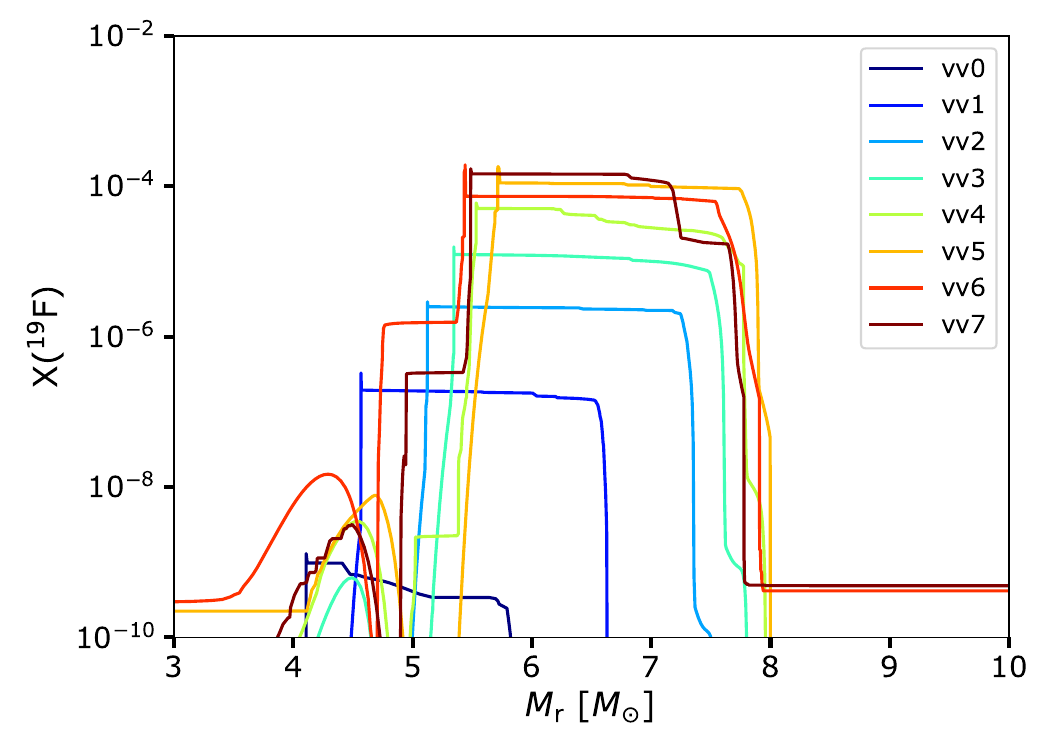}
\caption{
Pre-supernova abundance profiles of $^{14}$N (left panel) and $^{19}$F (right panel) predicted by our models.
}
\label{fig:presn}
\end{figure}

\subsection{Stellar yields, mass cut and dilution}

\begin{figure}[t]
\centering
\includegraphics[width=0.8\textwidth]{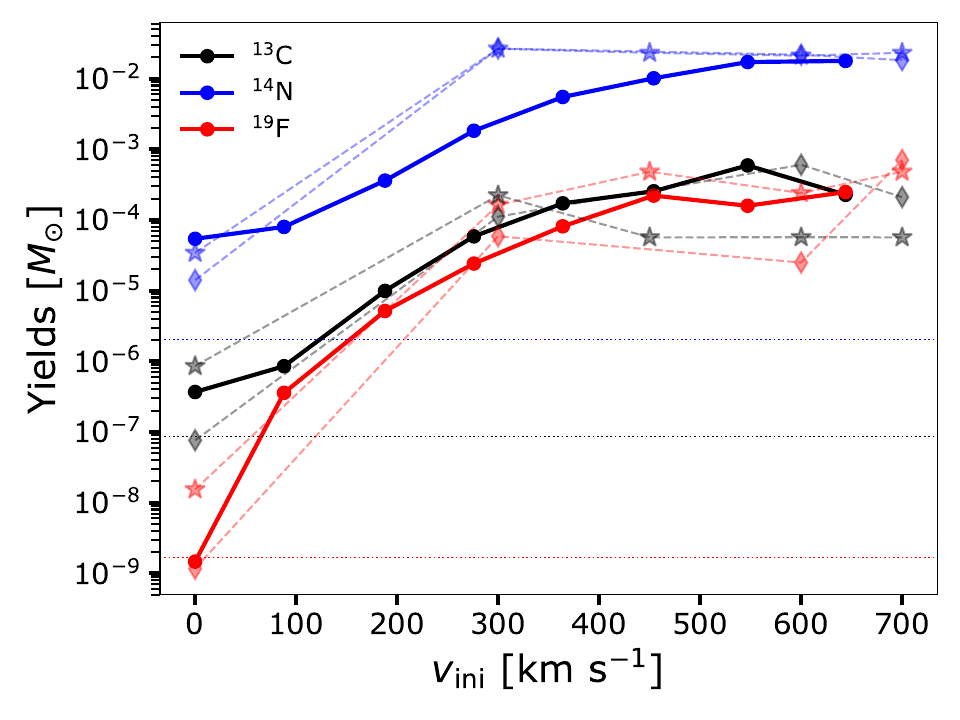}
\caption{Yields (in $M_{\odot}$) of $^{13}$C, $^{14}$N and $^{19}$F in the H- and He-rich regions, with the mass cut set at the top of the CO core (defined as the location where the $^{4}$He mass fraction falls below $10^{-2}$), shown as a function of initial velocity. The thin dotted horizontal lines represent the initial mass of these isotopes contained within the stars. The dashed patterns show the yields of the 15 (diamonds) and 25 $M_{\odot}$ models at [Fe/H]~$=-4$ (set E) of Roberti et al. (2024) \cite{roberti24}.}
\label{fig:yields}
\end{figure}

At the end of evolution, $^{14}$N and $^{19}$F (when present) are found at the base of the H-rich and He-rich layers, respectively. 
Beneath the H-rich layers, $^{14}$N is largely destroyed, contributing in part to the synthesis of $^{19}$F. 
The final abundances of $^{14}$N and $^{19}$F scale with the initial rotation (Fig.~\ref{fig:presn}), leading to steadily increasing yields at higher rotation rates (Fig.~\ref{fig:yields}).
The yields of $^{14}$N (blue) and $^{19}$F (red) show, as expected, nearly identical trends with initial rotation, highlighting the strong dependence of $^{19}$F synthesis on $^{14}$N. 
The yields of $^{13}$C, produced through the back-and-forth mixing process (Sect.~\ref{sec:backforth}), follows a similar pattern.

\citet{roberti24} investigated in detail the nucleosynthesis in 15 and 25~$M_{\odot}$ stars at [Fe/H]~$=-4$, $-5$, and zero metallicity, considering various initial rotation rates. The explosive yields of $^{13}$C, $^{14}$N, and $^{19}$F for the [Fe/H]~$=-4$ models (similar in metallicity to our models) are qualitatively similar to ours (Fig.~\ref{fig:yields}, dashed lines), particularly in the fast-rotation regime where the yields converge toward similar values. This indicates that the yields of these nuclides show only a modest dependence on the physical inputs, the treatment of rotation, and the explosive assumptions.

\begin{figure}[t]
\centering
\includegraphics[width=0.49\textwidth]{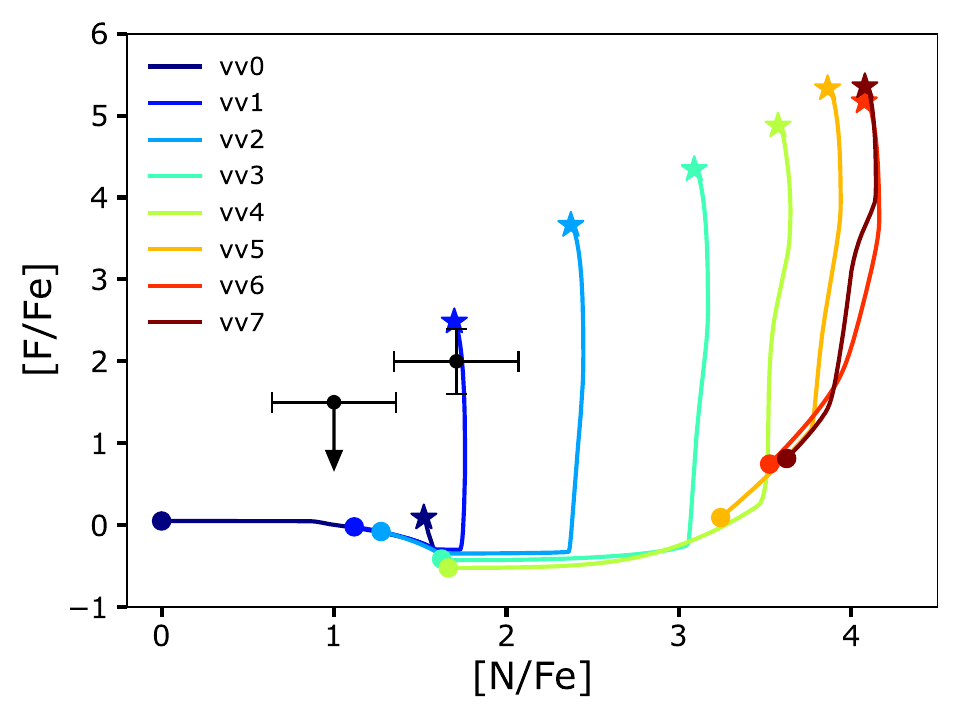}
\includegraphics[width=0.49\textwidth]{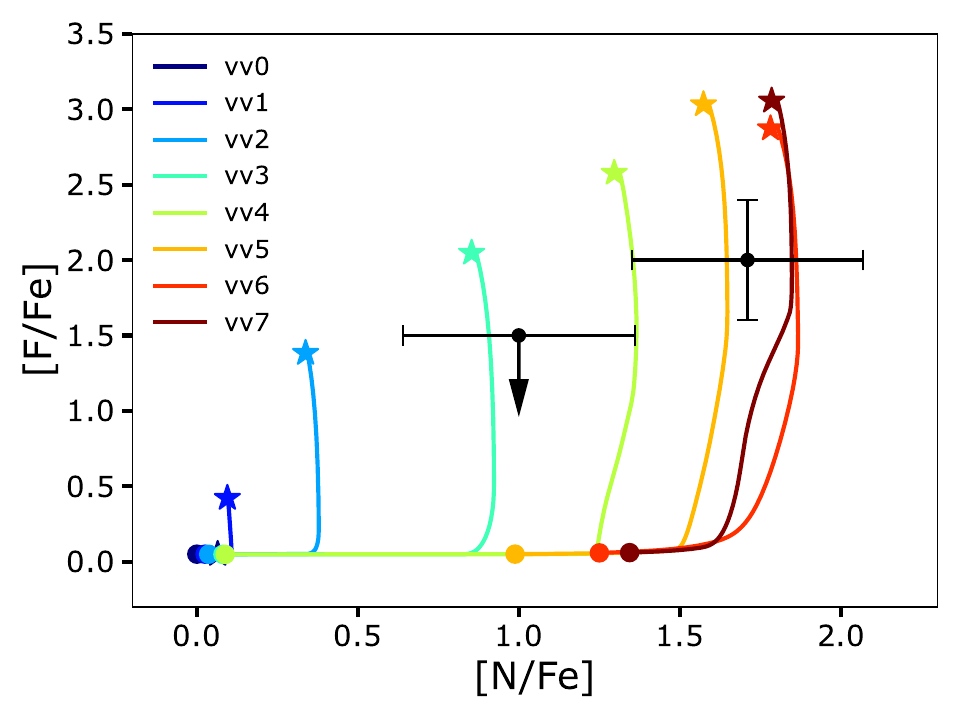}
\caption{
Predicted [N/Fe] versus [F/Fe] ratios as the mass cut is varied in our models, from the final mass (i.e., winds only, shown as colored circles) to the top of the CO core (shown as colored stars). A solid line connects all possible [N/Fe]–[F/Fe] combinations in the corresponding model ejecta. 
The left panel shows the case where no dilution with ISM is considered (i.e. pure massive star ejecta).
The right panel shows the case where the dilution factor (Eq.~\ref{eq:dil}) is fixed to $f_{\rm dil} = 0.995$.
Black symbols mark the two CEMP-no stars CS 29498$-$043 and CS 29502$-$092.
}
\label{fig:nfe}
\end{figure}

In Fig.~\ref{fig:yields}, the mass cut\footnote{At the time of the explosion, the mass cut defines the boundary between the material ejected and the material locked into the remnant.} is set at the top of the CO core, where the $^{4}$He mass fraction falls below $10^{-2}$. Since the exact location of the mass cut is uncertain, we varied it freely from the top of the CO core to the final stellar mass, the latter corresponding to pure wind ejecta. Overall, [N/Fe] and [F/Fe] ratios increase with progressively deeper mass cuts (Fig.~\ref{fig:nfe}, left panel). Because F-rich layers lie deeper than N-rich layers, [N/Fe] rises first, followed by [F/Fe]. 
It is worth noting that the non-rotating model produces some secondary nitrogen via the CNO cycle, reaching [N/Fe] values up to $\sim 1.5$ (dark blue star in Fig.~\ref{fig:nfe}, left panel). However, the non-rotating model does not produce any fluorine, regardless of the adopted mass cut (dark blue pattern in Fig.~\ref{fig:nfe}, left panel). As soon as some rotation is included (e.g. the vv1 model), some F is synthesized. 
Mixing the stellar ejecta with the $Z=10^{-5}$, $\alpha$-enhanced ISM (Sect.~\ref{sec2}) shifts the patterns toward [N/Fe]~$=0$ and [F/Fe]~$=0$ (Fig.~\ref{fig:nfe}, right panel).

\begin{table}[h]
\caption{Stellar parameters and abundances of the fluorine-rich CEMP-no star CS 29498$-$043.}\label{table:2}%
\begin{tabular}{@{}lllll@{}}
\toprule
                                    & CS 29498$-$043    & Ref.                              \\
\midrule
$T_{\rm eff}$              & 4440                         & \cite{roederer14a}   \\  
$\log(g)$                    & 0.5                            & \cite{roederer14a}  \\  
$[$Fe/H]                    & $-3.85$                    & \cite{roederer14a}   \\  
Binarity?                    & no                            & \cite{hansen16b}       \\  
\midrule
A(Li)                            &  $<-0.05$               & \cite{roederer14a}       \\
$[$C/Fe]                      & $2.72\pm0.25$      & \cite{roederer14a}          \\  
$[$N/Fe]                      & $1.71\pm0.36$      & \cite{roederer14a}        \\  
$[$O/Fe]                      & $2.37\pm0.15$     & \cite{roederer14a}         \\  
$[$F/Fe]                      & $2.00\pm0.4$      & \cite{mura25}          \\  
$[$Na/Fe]                    & $1.03\pm0.23$     & \cite{roederer14a}               \\  
$[$Mg/Fe]                   & $1.78\pm0.18$      & \cite{roederer14a}         \\  
$[$Al/Fe]                      & $0.75\pm0.25$    & \cite{roederer14a}          \\  
$[$Sr/Fe]                      & $0.08\pm0.24$    & \cite{roederer14a}         \\  
$[$Ba/Fe]                      & $-0.51\pm0.15$  & \cite{roederer14a}             \\  
$^{12}$C/$^{13}$C     & $8 \pm 3$             & \cite{roederer14c}     \\  
\bottomrule
\end{tabular}
\end{table}

\section{Comparison with the Fluorine-rich CEMP-no star CS 29498$-$043}\label{sec4}

Determining fluorine abundances in stars is challenging, which makes observational data extremely scarce. To our knowledge, fluorine has only been measured in the CEMP-no star CS 29498$-$043 \citep{mura25}. Upper limits were reported for three additional CEMP-no stars by \cite{mura25}. However, for two of them (HD 126587 and HE 1116-0634), the latest observational studies having reported carbon abundances find [C/Fe] ratios below the canonical $0.7$ threshold for CEMP-no classification\footnote{For HD 126587, \cite{roederer14a} report [C/Fe]~$=-0.01 \pm 0.25$; for HE 1116-0634, \cite{hollek11} report [C/Fe]~$=0.08 \pm 0.20$.}, suggesting that these stars are instead normal metal-poor stars. We therefore restrict our sample to CS~29498$-$043, whose properties and abundances are listed in Table~\ref{table:2}.

To match the observed chemical abundances of this star with our models, we follow a similar procedure as described in Sect.~6.2 of \citet{choplin21}. For each star, the best-fitting massive star nucleosynthesis model is identified by minimizing the $\chi^2$ value, which quantifies the difference between observed and theoretical abundances \cite[Eq.~7 in][]{choplin21}. This is done by mixing the material ejected by our massive star models with interstellar medium (ISM) material. During the fitting, the abundance of each isotope $i$ is computed as
\begin{equation} 
X_{\rm DIL}^{i} = (1-f_{\rm dil}) \, X_{\rm YIELDS}^{i} + f_{\rm dil} \, X_{\rm ISM}^{i} 
\label{eq:dil} 
\end{equation} 
where $X_{\rm YIELDS}^{i}$ and $X_{\rm ISM}^{i}$ are the mass fractions of isotope $i$ in the stellar ejecta and ISM (from Table~\ref{table:ism}), respectively. The dilution factor $f_{\rm dil}$, varied between 0 and 1, is adjusted to minimize $\chi^2$. It controls the proportion of massive star ejecta mixed into the ISM. 
The mass cut $M_{\rm cut}$ is set at the top of the CO core.

The reduced\footnote{The reduced $\chi^2$ is defined as $\chi_\nu^2 = \chi^2 / N_{\rm ab}$ where $N_{\rm ab}$ is the number of data points.} $\chi_\nu^2$ gets smaller and smaller with increasing initial rotation rate (Fig.~\ref{fig:fits}). 
CS~29498$-$043 is best fitted by the $v_{\rm ini}/v_{\rm crit} = 0.6$ model, which has the smallest $\chi_\nu^2$ value of 1.29. The mass cut is $M_{\rm cut} = 4.74 M_{\odot}$ with $f_{\rm dil} = 0.996$. It corresponds to a material made of $\sim 15 M_{\odot}$ of stellar ejecta diluted in $\sim 3.7 \times 10^3$~$M_{\odot}$ of ISM. 
The [C/Fe], [N/Fe], [O/Fe], [Na/Fe], [Mg/Fe], and [Al/Fe] ratios in CS~29498$-$043 can be reproduced by this model within the observational uncertainties (with Na and Al at the edge of the error bars; Fig.~\ref{fig:fits}, bottom panel). The predicted [F/Fe] ratio is [F/Fe]~$=2.8$, which is inconsistent with the observed value of [F/Fe]~$=2.0 \pm 0.4$. A lower [F/Fe] ratio can be achieved by assuming a higher dilution factor, but this worsens the fit for other elements. The best-fit $v_{\rm ini}/v_{\rm crit} = 0.2$ and 0.3 models can reproduce the [F/Fe] ratio but fail to correctly match other elements (e.g., [Mg/Fe] or [Al/Fe]). Considering the abundances in terms of\footnote{This quantity is defined as $\log_{10} \epsilon \, \mathrm{(X)} = \log_{10}(N_{\rm X}/N_{\rm H})_{\star} + 12$, with $N_{\rm X}$ and $N_{\rm H}$ the number densities of element X and hydrogen in the observed star.} $\log_{10} \epsilon \, \mathrm{(X)}$ or [X/H] ratios (instead of [X/Fe]) leads to similar results and the same conclusions.

\subsection{A possible solution to the overproduction of Fluorine}

The $^{15}$N($\alpha,\gamma$)$^{19}$F and $^{19}$F($\alpha,p$)$^{22}$Ne reaction rates have been identified as among the most uncertain and influential for fluorine nucleosynthesis in AGB stars \citep{cristallo14}. By varying these reaction rates individually by up to a factor of 100, \citet{cristallo14} found that the final surface abundance of $^{19}$F can be reduced by up to a factor of 10. 

Here, the rate of the $^{15}$N($\alpha,\gamma$)$^{19}$F reaction is taken from NACRE \citep{angulo99, xu13}. Other evaluations of this rate include \citet{caughlan88}, \citet{wilmes02}, and \citet{iliadis10a}. At astrophysically relevant temperatures, the recommended rates from these sources differ by at most 20\%, except for the original estimate by \citet{caughlan88}, which was approximately a factor of 50 higher. According to \citet{iliadis10a}, the uncertainty in their evaluated rate is roughly a factor of 4 within the temperature range $0.1 \lesssim T \lesssim 0.2$~GK (see also Fig.~9 in \citet{fang24}).

For the $^{19}$F($\alpha,p$)$^{22}$Ne reaction, we used the rate of \citet{caughlan88}. In the temperature range $0.1 \lesssim T \lesssim 0.3$~GK, the more recent recommended rate of \citet{ugalde08} is up to a factor of $\sim 4$ lower than that of \citet{caughlan88}. Nevertheless, uncertainties of approximately a factor of $\sim 10$ can be expected at these temperatures (see Fig.~9 in \citet{ugalde08}). The more recent determination by \citet{dagata18}, using the Trojan Horse method \citep{tumino21}, suggests an increased rate relative to \citet{ugalde08}, yielding a recommended rate close to the original \citet{caughlan88} estimate.

\begin{figure}[h!]
\centering
\includegraphics[width=0.8\textwidth]{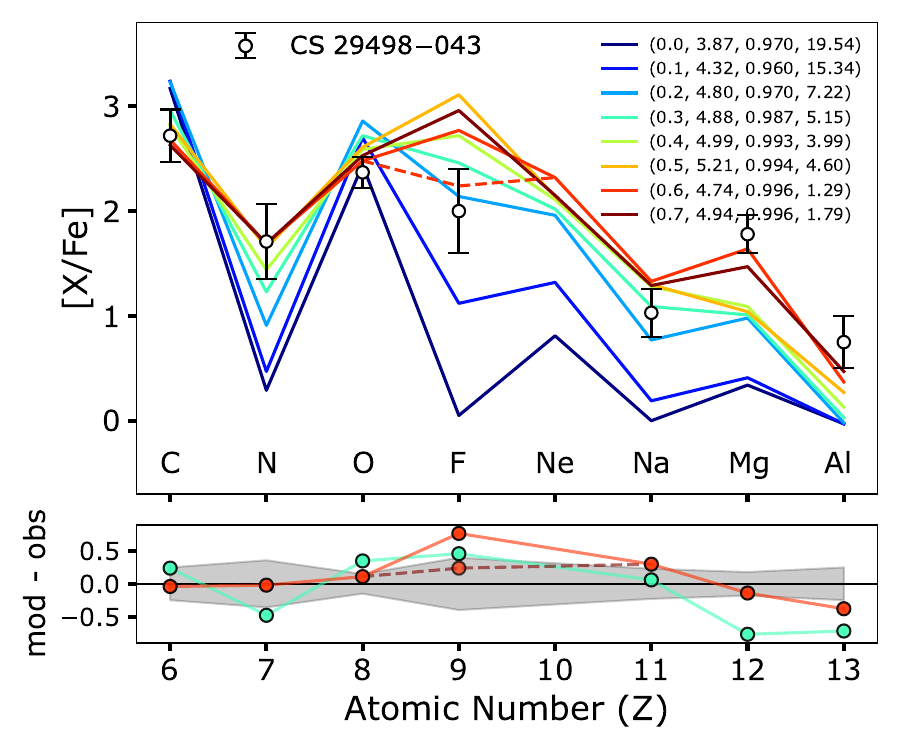}
\caption{
Best-fit results for our eight models for the fluorine-rich CEMP-no star CS 29498$-$043. Numbers in parentheses indicate the initial $v_{\rm ini}/v_{\rm crit}$, mass cut (in $M_{\odot}$), dilution factor (Eq.~\ref{eq:dil}), and reduced $\chi_\nu^2 = \chi^2 / N_{\rm ab}$ where $N_{\rm ab}$ is the number of data points. The dashed line corresponds to the $v_{\rm ini}/v_{\rm crit} = 0.6$ model with the $^{15}$N($\alpha,\gamma$)$^{19}$F rate reduced by a factor of 3 and the $^{19}$F($\alpha,p$)$^{22}$Ne rate increased by a factor of 5.  
The bottom panel shows the residuals of the fits for the $v_{\rm ini}/v_{\rm crit} = 0.3$ and 0.6 models, including the model with modified reaction rates (dotted line).
}
\label{fig:fits}
\end{figure}

We found that the fluorine abundance of CS~29498$-$043 can be reproduced by the $v_{\rm ini}/v_{\rm crit} = 0.6$ model if this model is recomputed while simultaneously decreasing the rate of $^{15}$N($\alpha,\gamma$)$^{19}$F by a factor of three and increasing the rate of $^{19}$F($\alpha,p$)$^{22}$Ne by a factor of five.  
Given the uncertainties in these reaction rates (see previous discussion), we argue that such variations remain within the acceptable range.  
The $v_{\rm ini}/v_{\rm crit} = 0.6$ model with modified rates produces nucleosynthesis very similar to the original model, except for a reduced fluorine yield. The resulting best-fit to CS~29498$-$043 now has [F/Fe]~$=2.2$ (instead of 2.8; Fig.~\ref{fig:fits}), which is consistent with the observed fluorine abundance.

\section{Discussion}\label{sec5}

\subsection{Origin of iron-peak elements}

In this work, the source star metallicity is $Z=10^{-5}$, and the H--He layers therefore already contain some Fe (and nearby elements) at levels similar to those observed in CS~29498$-$043. In the simplest scenario, this Fe originates from one or more previous massive, likely Pop.~III stars.  
Alternatively, some Fe may be ejected from massive stars experiencing extensive fallback if the mixing mechanism proposed by \citet{umeda02} is active (the mixing \& fallback scenario). In this case, the mass cut can lie well above the Fe core, but pre-SN mixing will transport some Fe into the ejecta.

It is also possible that, given an initial velocity dispersion, Fe is preferentially produced by slower rotators, which develop smaller CO cores (Table~\ref{table:1}), leading to more compact progenitors \citep{mapelli20} and deeper mass cuts. The high-velocity tail, corresponding to more compact massive stars that are less likely to explode, may be responsible for the formation of CEMP-no stars.

\subsection{Impact of initial mass and mass loss}

In more massive stars, mass loss can begin to play a role even at low metallicity, either because the star reaches the $\Omega\Gamma$ limit \cite{maeder00a} or because of self-enrichment, whereby the surface becomes metal-rich enough to trigger mass loss through radiatively line-driven winds \cite{hirschi07, meynet10, aryan24}. 
Reaching the $\Omega\Gamma$ limit occurs primarily in very massive stars (typically around 150 M$_\odot$ for Population III stars). The substantial mass loss that follows may even prevent the star from exploding as a pair-instability supernova \cite{Sylvia2008IAUS}. At present, however, the impact of this $\Omega\Gamma$-limit–driven mass loss on the enrichment of the interstellar medium with newly synthesized elements remains to be investigated.

Mass loss triggered by self-enrichment occurs, at the earliest, during the core He-burning phase \cite{meynet10}. The resulting mass loss can be significant, especially if the star evolves toward the red part of the HR diagram, provided that the usual scaling law relating mass-loss rates to surface metallicity applies -- a point that still requires confirmation from stellar-wind models. The expelled material is enriched in CNO elements and thus contributes to the chemical enrichment of the interstellar medium.
In the present work, the models considered do not exhibit this behaviour, and most of the mass loss occurs during the supernova explosion. However, even if such a strong mass-loss episode were present, the results would not change significantly. The material expelled through these winds would have nearly the same chemical composition as the corresponding stellar layers ejected during the supernova.

\subsection{Metallicity dependence of the results}

Reducing the metallicity from $Z = 10^{-5}$ to even lower, but still non-zero, values would certainly affect the yields for a given initial mass and rotation. However, the same qualitative mixing processes (characterized by repeated exchange between the H- and He-burning regions) would still operate, enabling the ejection of a CNO-enriched envelope (see, for example, the models at $Z = 10^{-5}$ and $10^{-8}$ in \citet{meynet10}).

Moving to higher metallicities (from $Z \simeq 10^{-4}$ upward) would lead to a substantial enrichment in s-process elements \cite{frischknecht16, choplin18, limongi18}, which is incompatible with the characteristics of CEMP-no stars. 
At even higher metallicities, the rotational mixing between the H- and He-burning shell still operates but becomes less efficient. The main physical reason is that higher metallicities lead to larger temperature contrasts between the H- and He-burning regions, increasing the spatial separation between these zones and lengthening the mixing timescale beyond the evolutionary timescale. 
Consequently, the scenario explored here only operates up to a limiting metallicity, currently estimated to lie between $10^{-5}$ and $10^{-4}$.

Finally, an interesting case is that of Population III stars. As discussed by \citet{roberti24} and \citet{tsiatsiou25}, Pop III models can be significant producers of fluorine. Explaining fluorine-rich CEMP-no stars through Population III progenitors requires some level of iron enrichment, which can be achieved by invoking faint supernova models that include mixing during the explosion and strong fallback \cite{Umeda2003}. 
We do not explore this topic further here. Let us simply note that, at present, Population III stars incorporating both rotation and mixing and fallback mechanisms remain promising candidates for the source stars of fluorine-rich CEMP-no systems.

\subsection{Model limitations and caveats}

There are additional limitations of the present work that we mention here without discussing in detail, as they still need to be explored through dedicated stellar models. First, we have adopted specific prescriptions for $D_{\rm shear}$ and $D_{\rm eff}$. Other choices are possible \cite[e.g.][]{meynet13, Nandal2024}, and these would in fact lead to different results. Moreover, the effects of an internal dynamo have not been included in this study (we considered non-magnetic models). Incorporating magnetic fields could significantly alter the outcomes.
We have also considered single-star evolution -- or, more precisely, stars that do not experience tidal interactions, mass transfer, or common-envelope episodes.

\section{Conclusions}\label{sec6}

In this work, we investigated whether the chemical composition of the first detected fluorine-rich CEMP-no star, CS~29498$-$043, can be explained by the H/He-rich layers of a previous very low-metallicity ($Z=10^{-5}$) 20~$M_{\odot}$ star with initial rotational velocities in the range $0 < v_{\rm ini}/v_{\rm crit} < 0.7$ ($0 < v_{\rm ini} < 644$~km~s$^{-1}$). 

While non-rotating very low-metallicity massive star models produce no fluorine in their H/He layers, including rotation allows efficient fluorine synthesis in the He-burning layers (as previously found), at levels comparable to those observed in CS~29498$-$043 (Figs.~\ref{fig:nfe} and \ref{fig:fits}).  
The best fit is obtained for the 20~$M_{\odot}$ model with $v_{\rm ini}/v_{\rm crit} = 0.6$ ($v_{\rm ini} = 547$~km~s$^{-1}$), which reproduces the C, N, O, Na, Mg, and Al abundances of CS~29498$-$043 within observational uncertainties, but predicts [F/Fe]~$=2.8$, higher than the observed value of [F/Fe]~$=2.0 \pm 0.4$.  
By simultaneously varying the $^{15}$N($\alpha,\gamma$)$^{19}$F and $^{19}$F($\alpha,p$)$^{22}$Ne reaction rates within their acceptable ranges, the predicted [F/Fe] can be reduced to 2.2, resulting in an overall abundance pattern compatible with CS~29498$-$043.  
Extremely metal-poor, rotating massive stars may thus be viable progenitors for explaining fluorine-rich CEMP-no stars.

Because fluorine-rich layers lie directly beneath the nitrogen-rich layers in the stellar structure, F-rich ejecta are necessarily N-rich, whereas N-rich ejecta may or may not be F-rich. This asymmetry is illustrated in Fig.~\ref{fig:nfe}, where models can reach high [N/Fe] with low [F/Fe], but not the reverse. Consequently, this scenario for CEMP-no stars is valid only if CEMP-no stars are not simultaneously F-rich and N-poor. For example, a CEMP-no star with [N/Fe]~$\simeq 0$ and [F/Fe]~$\gtrsim 1$ would be incompatible with this scenario.

A potentially interesting test of this scenario is the expected correlation between nitrogen and fluorine, which would require measurements of both abundances in a sample of CEMP-no stars.  
AGB stars could also produce N and F simultaneously, but they likely had not contributed at these early times.  
While our models do not experience proton ingestion events, such phenomena have been shown to occur frequently in zero- or very low-metallicity massive stars and to be a source of primary nitrogen \citep{heger10, clarkson20, roberti24}. However, fluorine may not be produced during these events, as reported by \citet[][Sect.~5.2]{roberti24}. Finally, fluorine potentially produced during the explosion, either via the $\nu$-process \citep{woosley88} or during explosive H-burning \citep{mura25}, is not accompanied by nitrogen production.  
An observed N--F correlation could therefore indicate the operation of rotational mixing in early massive stars. Nevertheless, alternative scenarios, such as proton ingestion during evolution followed by fluorine production during the supernova explosion, cannot be excluded.

\backmatter

\bmhead{Acknowledgements}

A.C. is post-doctorate F.R.S-FNRS fellow.

%%===========================================================================================%%
%% If you are submitting to one of the Nature Portfolio journals, using the eJP submission   %%
%% system, please include the references within the manuscript file itself. You may do this  %%
%% by copying the reference list from your .bbl file, paste it into the main manuscript .tex %%
%% file, and delete the associated \verb+\bibliography+ commands.                            %%
%%===========================================================================================%%

\bibliography{astro.bib}

\clearpage
\newpage

\end{document}